# MOLS 2.0: Software Package for Peptide Modelling and Protein-Ligand Docking

D. Sam Paul and N. Gautham

Centre of Advanced Study in Crystallography and Biophysics,
University of Madras, Chennai 600025. India.

**Abstract**

We have earlier developed an algorithm to perform conformational searches of proteins and peptides, and to perform docking of ligands to protein receptors. In order to identify optimal conformations and docked poses, this algorithm uses mutually orthogonal Latin squares (MOLS) for rationally sampling the vast conformational (or docking) space, and then analyses this relatively small sample using a variant of mean field theory. The conformational search part of the algorithm was programmed as MOLS 1.0. The docking portion of the algorithm, which allows only 'flexible ligand – rigid receptor' docking, was programmed as MOLSDOCK. Both are FORTRAN based command-line-only molecular docking computer programs, though a GUI was developed later for MOLS 1.0. Both the conformational search and the 'rigid receptor' docking parts of the algorithm have been extensively validated. We have now further enhanced the capabilities of the program by incorporating 'induced fit' side-chain receptor flexibility for docking peptide ligands. Benchmarking and extensive testing is now being carried out for the 'flexible receptor' portion of the docking. Additionally, to make both the peptide conformational search and docking algorithms (the latter including both 'flexible ligand – rigid receptor' and 'flexible ligand – flexible receptor' techniques) more accessible to the research community, we have developed MOLS 2.0, which incorporates a new Java-based Graphical User Interface (GUI). Here we give a detailed description of MOLS 2.0. The source code and binary of MOLS 2.0 are distributed free (under GNU Lesser General Public License) for the scientific community. They are freely available for download at **https://sourceforge.net/projects/mols2-0/files/.**

**Introduction**

The MOLS technique has been developed in our laboratory to address the issue of combinatorial explosion that occurs in computational searches for optimal protein and peptide structure [1]. Using mutually orthogonal Latin squares, the algorithm systematically identifies a small but completely representative sample of the multidimensional search space. The energy values are calculated at each of the sampled points. These energy values are then analyzed using a variant of the mean field method [2] to obtain the minimum energy conformation [3]. We later extended the technique to search for the optimal conformation, position and orientation of a small molecule ligand on the receptor [4, 5]. This latter problem is crucial in drug design [6]. The extension of the technique was straightforward. The search space is now considered to include, along with the conformation, the position and orientation of the ligand on the receptor surface. The energy function includes not only the conformation energy of the ligand, but also the interaction energy between the ligand and the receptor [4]. Application of the MOLS method then yields the optimal pose and conformation simultaneously.

The computer programs that executed the above procedure were lacking in many features, and this made them less accessible to the general users. They relied on command line operations, demanded expertise in the Linux operating system, and were not packaged together as one unit. Thus, one had to use separate programs to build the model, to carry out the MOLS sampling, and to analyze and display the results. An earlier attempt [7] to automate some of these functions resulted in an application that was suitable only when using peptides as ligands.

Another drawback, particularly of the docking algorithm was that only the ligand was allowed to change conformation on binding, while the receptor was held rigid. However, in the living cell, receptor proteins are not rigid, but may flex and move to accommodate the ligand [8–10]. This becomes clear when we compare the X-ray structure of a receptor in complex with its cognate ligand, with the structure of the same receptor without the ligand. Gutteridge and Thornton [11] studied conformational changes that occur on ligand binding in the case of 60 enzymes. By superposing the *apo* and *holo* forms of each enzyme, they observed that about 75% of the enzymes have C$\alpha$ RMSD less than 1 Å between the two forms; 91% of the enzymes have C$\alpha$ RMSD less than 2 Å. Thus structural changes in the protein as a whole are not large [12]. However, other studies showed that significant differences in side-chain conformation occurs in 60% of the proteins between the *holo* and *apo* forms upon ligand binding [13, 14], though main-chain conformations are largely preserved. These results imply that while designing computational docking algorithms, it is necessary to allow for receptor flexibility, especially side-chain flexibility, as much as for the conformational changes in the ligand. Recently, many docking tools [15–21] have incorporated different techniques for receptor flexibility.

To address these issues, here we report MOLS 2.0 as a complete program package, with a convenient Java-based GUI. The GUI allows ease of input of data, ease of viewing the results, as well to access to both the conformational search program as well as the docking program. We have used Jmol: an open-source Java viewer for chemical structures in 3D (http://www.jmol.org/) in our GUI. Many new features have been incorporated. The chief among these is a 'molecule-builder' unit that allows the user to build and input new organic ligand molecules graphically. Another important new feature is the introduction of 'flexible receptor docking'

**Methods**

**The MOLS algorithm**

The MOLS method developed in our laboratory searches the potential energy surface in an apparently exhaustive manner to locate all the low energy conformers. This method is based on the technique, taken from the field of experimental design, of using mutually orthogonal Latin squares (MOLS) to sample the experimental space, in this case the conformational space [1, 22]. In experimental design MOLS are used to systematically sample the space of the variables. MOLS allows the experimenter to finish the experiment with a relatively small number of runs ($M^2$) instead of examining all possible combinations of values ($M^N$) of the variables, where $N$ is the number of variables and $M$ is the number of possible values of each variable. The MOLS method as applied to conformational search, and to docking has been described in detail elsewhere [1, 22–24]. Here only a brief description is given for completeness. For simplicity, we will first consider the method as applied to the prediction of the minimum energy structure of a peptide. The 'experimental space' in this case is the conformation space of a peptide, which may be defined as the set of all possible combinations of all values of all its variable torsion angles. Thus, if there are $n$ such torsion angles, each which can take up $m$ different values, there are $m^n$ different combinations of these values. In other words, there are $m^n$ different conformations for the peptide. For each conformation a potential energy may be calculated, and therefore the multi-dimensional conformational space can also be considered to be potential energy surface of the peptide. The task then in peptide structure prediction is to locate the minimum value on this surface, i.e. the minimum energy conformation. (As we note from our studies [24], and from those of others [25, 26], there is not just a single, clearly identifiable minimum in the conformational energy space. There are, instead, a large number of different conformations of approximately equally low energy, any one of which may be the appropriate conformation in a given chemical or biological context). A brute force search for low energy conformations will lead to combinatorial explosion and, for peptides longer than about 15 residues, is beyond the capabilities of even the most powerful current-day computers. There are several methods to make the search more tractable [27–30]. The MOLS method is one of them. Here, we use mutually orthogonal Latin squares [31] to systematically choose a set of $m^2$ points (or conformations) from the conformational space of the peptide. Mutually orthogonal Latin squares are arrangements of the values of the variables such that all possible pairwise combination of the values are represented in the sample, without any point being repeated. Such an arrangement allows the choice of $m^2$ combinations of the values of the variables such that each combination represents one conformation of the molecule. The procedure will thus specify $m^2$ conformations. The potential energy for each conformation is calculated. The $m^2$ energy values so obtained are then analyzed using a variant of the mean field technique [2] to arrive at the conformation with the lowest energy. We note here that there are a very large number of ways in which we can choose the sets of mutually orthogonal Latin squares, and hence the $m^2$ points in conformational space. Each choice, after the analysis, can lead to either the same, or to a different low energy structure. We have shown [23] that repeating the calculations 1500 times (i.e. with $m^2$ points chosen, and a minimum energy structure identified, 1500 different times) is sufficient to sample the entire conformational space and identify the complete set of low energy conformations. It is easy to see that, given an appropriate potential energy function, and given a set of torsion angles that may be rotated to obtain all possible conformations, the MOLS method is suitable not only for peptides but for all organic molecules, many of which may dock at receptor sites and act as drugs.

The extension of the method to address the docking problem was straightforward [4]. The peptide or organic molecule is considered the flexible ligand, binding to a rigid receptor protein. Thus, in addition to the variable torsion angles of the ligand, we included, in the search space, the variables for the position and orientation of the ligand with respect to the protein. The energy function is also modified to include the binding energy between protein and ligand. The total energy is now therefore a function of the variable torsion angles of the peptide ligand, as well as its 'pose'. Again, MOLS is used to identify $m^2$ points in the enhanced search space. The value of the modified energy function is calculated at each of the identified points. These $m^2$ values are analyzed to arrive at the best conformation for the ligand as well as, simultaneously, its best pose at the receptor site. The method is suitable for all organic small molecules, including peptides, as ligands. [5, 32].

The peptide conformational search algorithm has been extensively tested [1, 24, 33]. Likewise, the docking algorithm, dubbed MOLSDOCK, has also been tested with 56 protein-peptide complexes [4], 45 protein-organic small molecule complexes [5]. In 73% of the protein-peptide test cases, the solution with the least RMSD as compared to the crystal structure was in the top 10% of energy when all the solutions were ranked in terms of energy. For 9 test cases, solutions with the lowest RMSD were also the ones with the lowest energy. In the case of the 45 protein-small organic molecule complexes, in 17 of them the solution with the least RMSD as compared to the crystal structure was in the top 10% of energy when all the solutions were ranked in terms of energy. For 2 test cases, solutions with the lowest RMSD were also the ones with the lowest energy. In both protein-peptide and protein-ligand docking, MOLSDOCK was capable of identifying alternate binding modes.

**<u>Receptor flexibility in MOLSDOCK</u>**

Keeping the receptor rigid during docking may be effective in some cases [21]. But when ligands bind, proteins often adopt different conformation, as proposed by the 'induced fit' theory [8]. Conformational changes in ligand-bound proteins are either seen in the backbone or as very small side-chain fluctuations [34]. Docking tools like AutoDock [18], AutoDockFR [19], GOLD [17], GLIDE [21] and RosettaLigand [16] allow receptor flexibility. Receptors have been treated as flexible by using techniques like soft docking, rotameric exploration, ensemble docking (or) multiple protein structures, molecular dynamics [35] and induced fit docking [36]. In order to include receptor flexibility into the MOLS docking algorithm, the search space is expanded to consider also the variable torsion angles of the receptor protein. The energy function now includes intra-protein energy (calculated as a function of the variable torsion angles of the protein), besides the intra-ligand energy and the protein-ligand interaction energy. (In the current version of the algorithm, receptor flexibility has been included only for peptide-protein docking). Since previous studies show little or no movement of the backbone atoms of a protein residue upon ligand binding [13], we decided to select only the residues that line the active site and make their side-chains alone to be flexible. The receptor backbone is kept rigid. Further, the 'minimal rotation hypothesis' [14] suggests that small motions in protein side-chains are sufficient for good docking [37]. Therefore, in MOLSDOCK the side-chain torsion angles of the flexible residues fluctuate only by up to ±15° from their position in the crystal structure. Next section gives a clear insight about the 'induced fit' docking in MOLSDOCK.

**'Induced fit' docking in MOLSDOCK**

While upgrading MOLSDOCK from 'rigid receptor flexible ligand' docking method to 'flexible receptor flexible ligand' docking method, we incorporated two major changes. Firstly, the search space is expanded to include the flexible residues of the receptor protein. Secondly, intra-protein energy, which assesses the receptor protein's conformation is added to the scoring function.

The following description will give more information about the variable parameters used by the method to predict the optimal protein-peptide complex. If the conformation of the peptide ligand is specified by 'm' torsion angles ($\theta_r$, r = 1, m) then, for defining the binding site of the peptide on the receptor, six parameters which describe the peptide's pose, i.e. three for the position and three for the orientation, have to be added making a total of m + 6 dimensions in the search space ($\theta_1$ to $\theta_{m+6}$). Since the receptor is also flexible, the side-chain torsion angles of the flexible residues of the receptor protein are also added. If the conformation of the flexible residues of the receptor is specified by 'n' torsion angles then it makes a total of m+6+n dimensions in the search space ($\theta_r$, r = 1, m+6+n). If each dimension is sampled at $N$ intervals, then the volume of the search space is $(N)^{m+6+n}$. The MOLS technique calculates the values of the scoring function at $(N)^2$ points in this space, and analyses them using a variant of the mean field technique [2], to simultaneously identify the optimum conformation of the peptide, its pose, and also the conformation of the side-chains of the protein flexible residues. The search space is defined on a discrete grid of *m+6+n* dimensions, therefore for each cycle of calculations the method identifies an optimum point on the grid. However the actual optimum may lie close to but not actually on the grid. Therefore the final step in identifying is to perform a gradient minimization to find the nearest off-grid optimum.

Here, the two systems involved in the docking are the receptor protein and the peptide ligand. The peptide sequence and the 3-dimensional structure of the receptor protein in .pdb format are given as input. For cases where the binding site is known, the coordinates of the centroid of the binding site of the peptide in the receptor protein are given. If the binding site is not known, Fpocket 2.0 may be used to find the binding site. If the flexible residues of the receptor protein are known they may be predefined. If the flexible residues are not known then, automatically, all the protein residues whose atoms are at a distance of less than 4.0 Å from any atom of the extended conformation of the ligand placed in the binding site are selected for receptor flexibility. Using the peptide sequence, the extended conformation of the peptide is built as an all-atom model, including all the hydrogens. All the bond lengths and bond angles are fixed at their equilibrium values [38], with the peptide bond torsion angle fixed at 180°. All the backbone torsion angles ($\varphi,\psi$) and side-chain torsion angles ($\chi$) of the peptide are taken as independent variables, and are allowed to vary. The centroid of the binding site is taken as the grid centre. The binding site in the receptor protein is defined by a 5.0 Å cubic grid box constructed around the centroid. The peptide ligand is placed at the grid centre. During the search, the peptide is allowed to move within the grid box with the peptide centre translating along the grid in steps of 0.14 Å. The receptor protein is protonated [39]. The side-chain torsion angles of the identified flexible residues are allowed to vary.

The range for each torsion angle of the peptide ligand is 0° - 360° with a search step size of 10°. Three parameters are used to represent the orientation of the ligand inside the cubic box. Two of these represent the position of a rotation axis, and the remaining one is the angle of rotation about this axis. To define the position of the axis, an imaginary unit sphere is constructed around the center of the cubic box. The axis is a line from the centre to a point on the surface of the sphere. The two spherical coordinates specifying the point on the surface are then used to represent the position of the axis of rotation. The range of the polar angle is from 0° to 180°. The range of the azimuthal angle is 0° to 180°. The range of the angle of rotation about the axis is 0° to 360°. All three angles specifying the orientation of the ligand are sampled in steps of 10°. The three translation parameters along the *x*, *y* and *z* axes each have a range of 2.5 Å on either side of the center, and the centroid of the ligand is moved in steps of 0.14 Å. The range for the side-chain torsion angles of the flexible residues in the receptor protein is ±15˚ from their position in the crystal structure. Only the centroid of the ligand is displaced within the defined cubic box during the docking simulation.

The scoring function is the weighted sum of three terms: the intra-peptide energy calculated using AMBER94 force field [40], the intermolecular interaction energy calculated using the PLP scoring function [41] and the intra-protein energy calculated using AMBER94 force field [40]. The AMBER force field is reported in units of Kcal/mole [40], whereas the PLP force field is reported in dimensionless units [41]. The total potential energy is reported in dimensionless units.

**Description of MOLS 2.0 software package**

MOLS 2.0 (Fig. 1) is equipped with MOLS - a conformation search tool for peptides, MOLSDOCK - a docking tool, Jmol viewer and a Molecule Builder. MOLSDOCK is equipped also with induced fit docking but currently only for peptide ligands. The Molecule Builder may be used to create drug candidates for docking.

**Prerequisites**

MOLS 2.0 requires Java 7. Fpocket 2.0 [42], available at http://fpocket.sourceforge.net/, is used for protein cavity detection in MOLS 2.0. Open Babel [43], is used for file format conversions and energy calculations inside the docking tool. Fpocket 2.0 and Open Babel 2.3.2 are shipped, and will be installed, along with MOLS 2.0. In case the basic docking algorithm, which is written in FORTRAN, needs recompilation, it must be carried out using 'Intel® Fortran Composer XE for Linux (http://software.intel.com/en-us/non-commercial-software-development). MOLS 2.0 is developed for Linux. The opening screen asks for the project name and directory. It provides access to the different tools.

**MOLS: Peptide structure prediction tool**

This tool takes an amino acid sequence as an input and generates for it a specified number of low energy structures. The sequence is specified using the single-letter amino acid code. Other options include controls to restrict the flexibility to the backbone alone, or to search over the entire conformational space of the molecule. There is also a '*rotamer*' search option, in which the side-chain dihedral angles are sampled from values given in a rotamer library [44]. Further options allow the generated structures to be minimized using conjugate gradient minimization, and to be clustered using *K-means* or *Hierarchical* clustering algorithms. Two options are provided for energy calculations – the AMBER94 force field [40] and ECEPP/3 force field [45].

**MOLSDOCK: Protein-ligand docking tool**

Inputs to the docking tool include, the ligand specifications, including structure, the receptor specifications and structure, the number of structures to be generated, and specifications of the binding site. The ligand, which could be any organic chemical compound, is specified in MDL Molfile (.mol) format. If such a description is not already available, the Molecule Builder in MOLS 2.0 may be used. The receptor, a protein, to which the ligand is to be docked, is specified in .pdb format. In the current version of the program only small ligand molecules having less than 30 rotatable bonds are allowed for docking. If the ligand is a peptide, a choice is offered between rigid-receptor docking, and induced-fit, i.e. flexible receptor docking.

Usually in molecular docking, except the protein and the ligand, no additional data need be given. However, if the binding site is known, for example from an experimentally determined structure of a protein-ligand complex, the *manual* button can be used, and the binding site can be defined by the coordinates of the centre of a box and its dimensions. If the binding site on the receptor protein is not known, the *auto* option can be used. Fpocket [42] is then used to select the best binding pocket. In the case of flexible receptor docking, if the flexible residues are known, they may be specified using the *Manual* option. Else, the *Auto* option will select the flexible residues.

The scoring function (or the objective function in the MOLS search) is the sum of the intra-molecule ligand energy, calculated using MMFF94 force field [46], and the intermolecular interaction energy between the protein and the ligand, calculated using the PLP scoring function [41].

**Test cases**

The sub-programs MOLS (Peptide Structure Prediction Tool) and MOLSDOCK (Protein – Ligand Docking Tool) of the MOLS 2.0 package have been already extensively tested [4, 5, 24, 32]. Some of these were re-run using the new programme package, with identical results. 'Induced fit' docking alone has been newly added to MOLSDOCK. Though tests of the conformational search and the 'rigid receptor' docking portion of the algorithm have been performed and reported earlier [4, 5], such benchmarking and extensive testing is only now being carried out for the 'flexible receptor' portion of the docking. Extensive tests of docking are being carried out on a number of test cases. A complete report of these tests will be presented in the future. In the meantime, users should take the results of the 'induced fit' docking with caution. Here, in order to indicate the possible results that may be obtained with MOLS 2.0, we report the results from one test case.

The peptide-protein complex, Gly-Ala-Trp complexed with γ-chymotrypsin [47], has been solved to a resolution of 1.6 Å (PDB ID: 8GCH). This structure was used as the initial test case. A model of the GAW tripeptide was first generated in the BUILDER subroutine. The peptide was docked to the structure of the unbound form of γ-chymotrypsin (PDB ID: 4CHA). Since the binding site of the ligand was known from the crystal structure, this was input to the program. Induced fit docking was carried out with the option for selecting the flexible residues in the protein set to *auto*. The program automatically identified 37 residues in the receptor protein whose structures could change on ligand binding. A total of 1500 docked structures were generated. For comparison, rigid docking was also carried out, using the same module MOLSDOCK. Again, 1,500 structures were generated. These structures were compared with the structure of the complex (PDB ID: 8GCH). The docking simulation was run on 2.4 GHz AMD Opteron Processor under the Linux operating system. The computational details are given in Table 1. The conjugate gradient minimization consumes the largest part (~75%) of the computation time.

**Results and discussions**

Table 2. summarizes the results obtained. In the following discussion, we specifically pick two structures from the total 1,500 predictions for the test case. Firstly, we identify the best sampled structure, i.e. the prediction that has the lowest backbone RMSD with respect to the native structure. The other structure we consider for discussions is the top ranked prediction, i.e. the prediction with the lowest total energy of all 1,500 predictions. For our analysis, we consider the crystal peptide structure in 8GCH.pdb as the native structure. The predicted structure of the peptide and the protein is superposed on the native structure to obtain the best overall RMSD. The backbone RMSD is then calculated for the peptide alone without any further rotation or translation. This procedure measures the differences not only in the structure of the peptide, but also in its pose with respect to the protein. Chung and Subbiah [48] have suggested that structures predicted within 2.00 Å backbone RMSD are in good agreement with the respective native counterparts. Table 2 shows that the top ranked structure of the 'induced fit' docking protocol has peptide backbone RMSD of 1.69 Å, which according to the criteria of Chung and Subbiah [48], is in good agreement with the native structure. This is as compared to the peptide backbone RMSD of 2.66 Å for the top ranked structure from 'rigid' docking. The backbone RMSD of the best sampled peptide structure (i.e. the predicted structure which is the closest in conformation to the native structure) in 'induced fit' and 'rigid' docking is 1.62 Å and 1.61 Å respectively, indicating that MOLS sampling is efficient in both 'induced fit' and 'rigid' docking. The best sampled structure in both 'induced fit' and 'rigid' docking is ranked within the top 5% of the energy ranked structures.

The position of the peptide ligand in the crystal structure of Gly-Ala-Trp complexed with γ-chymotrypsin [47] is the average of a structure in which the ligand is covalently bound to the receptor, and one in which it is not so bonded. In the crystal structure, the COOH terminal carbon of the ligand shows short contact distances to the oxygen atom of Ser195. Since docking techniques are usually developed for only non-bonded interactions, short contacts will be mostly filtered out during the course of docking. Therefore docking programs find the prediction of complex, such as this, a difficult task.

The top ranked peptide-protein complex structure in 'induced fit' docking has a total of 6 hydrogen bonds, of which 3 hydrogen bonds were also seen in experimental structure of the peptide-protein complex (Fig. 2). The top ranked peptide–protein complex structure in 'rigid' docking has 5 hydrogen bonds of which 2 were also seen in the experimental structure. (Fig. 3). However the energy values for the lowest energy structures in both models of docking are less than the value calculated using the same energy expression for the experimental structure. This may be due to the larger number of non-bonded contacts in the rigid docked model (101) and in the induced fit model (112) than in experimental structure (72).

Relative displacement error [49] is a metric used to rank different conformations and poses of a ligand docked to the receptor with respect to the known correct solution, which is the crystal structure. We calculated relative displacement error (abbreviated as RDE in Table 2) for the best sampled (i.e. the structure with least RMSD with respect to the experimental structure) and the lowest energy structures. The relative displacement error (RDE) of the lowest energy structure in 'rigid' and 'induced fit' docking is 55.36% and 53.11%. For the best sampled structure the RDE values are 43.96% for rigid docking and 46.78% for induced fit docking. These results indicate that, at least for this example, induced fit docking does not perform substantially better than rigid docking.

## **Conclusion**

MOLS 2.0 is a Java-based program that allows the use of useful tools developed in our laboratory, all based on the mutually orthogonal Latin squares sampling technique developed by us. The tools are: MOLS – a peptide conformational search tool to identify low energy peptide structures; MOLSDOCK – 'rigid protein – flexible ligand' docking tool. We have incorporated receptor flexibility in MOLSDOCK, therefore it could be used for 'induced fit docking', but currently only for docking peptide ligands. MOLS 2.0 also has Jmol molecule viewer and a built-in molecule builder, which may be used to create chemical compounds to be used for docking. In the one test case reported here 'induced fit' MOLSDOCK is better than rigid docking by one metric, but not by another. We are currently validating and optimizing 'induced fit' MOLSDOCK, by comparing it with other 'induced fit' docking tools, to increase its docking accuracy and scoring efficiency.

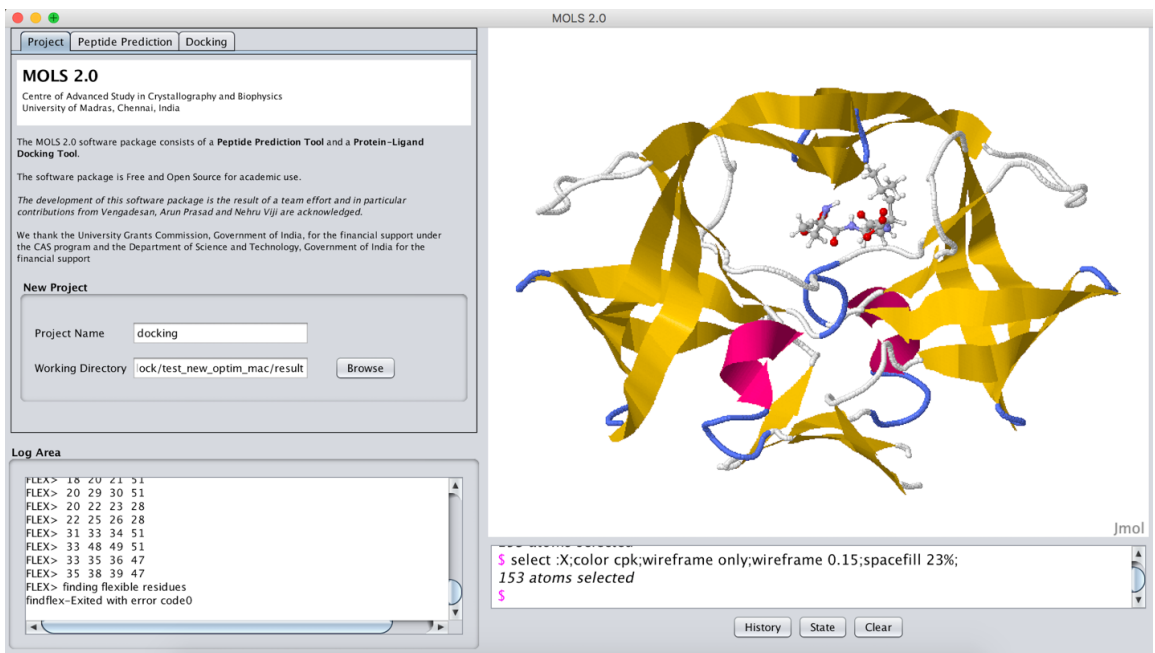

**Fig. 1** Graphical User Interface of MOLS 2.0

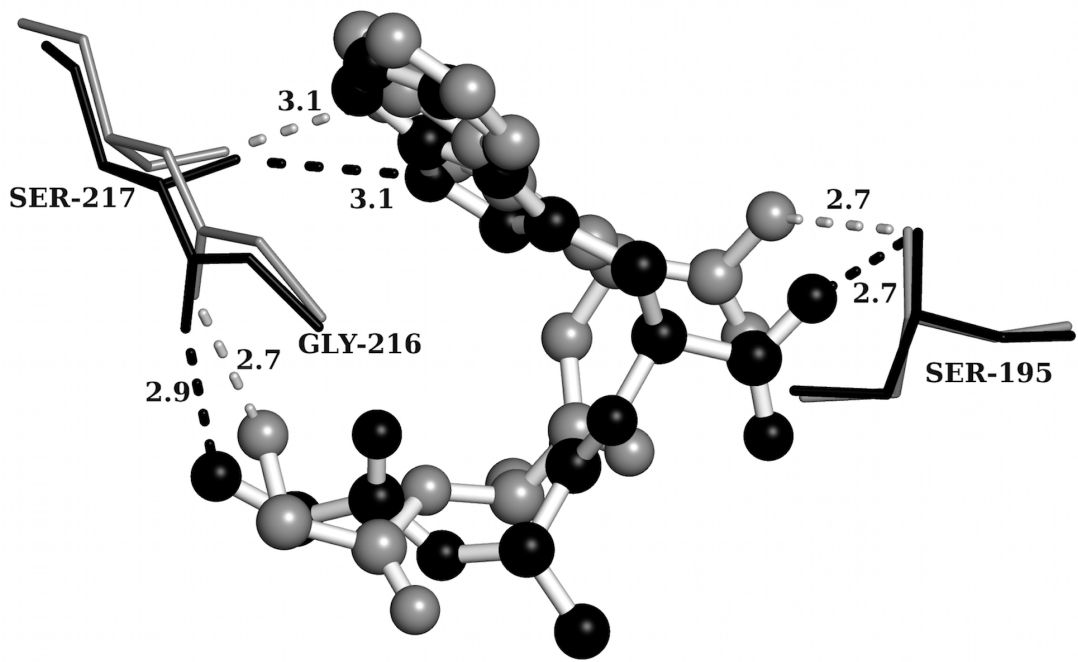

**Fig. 2** Conserved hydrogen bonds in 'induced fit' docking: Lowest energy peptide from induced fit docking superposed on the native. The three conserved hydrogen bonds between the native and the lowest energy prediction are shown. The lowest energy peptide and the native peptide are shown in gray and black respectively. The bond distances are in Ångstrom units.

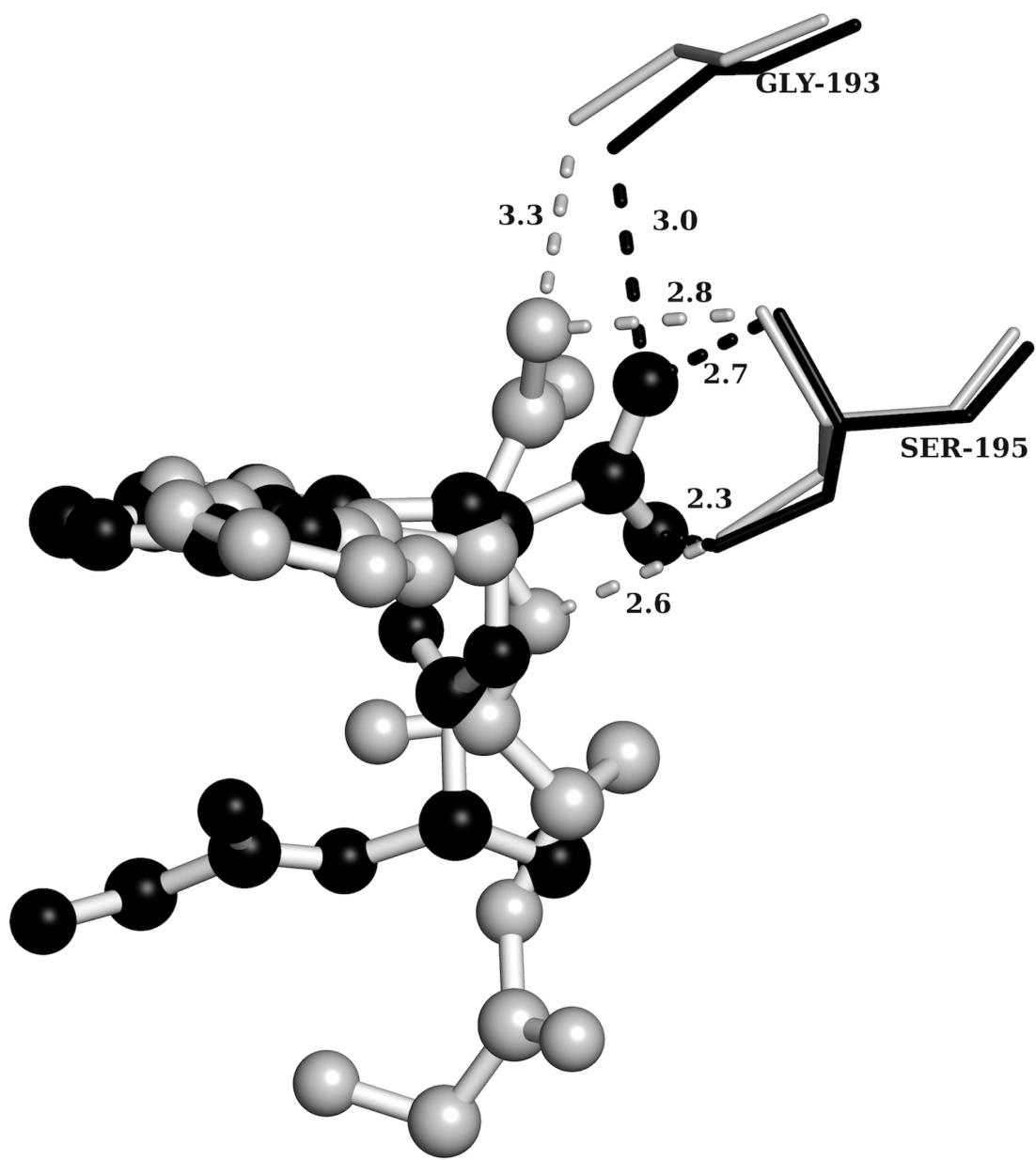

**Fig. 3** Conserved hydrogen bonds in 'rigid receptor' docking: Lowest energy peptide from rigid docking is superposed on the native. The two conserved hydrogen bonds between the native and the lowest energy prediction are shown. The lowest energy peptide and the native structure are shown in gray and black respectively. The bond distances are in Ångstrom units.

**Table 1.** Computation details of 'rigid' and 'induced fit' protein-peptide docking

|  | **Flexible residues** | **Total number of parameters**[a] | **cpu time (100 structures)** |
|---|---|---|---|
| **Rigid docking** | 0 | 14 | 10 m 15 s |
| **Induced fit docking** | 37 | 55 | 133 h 20 m |

[a] Total parameters include torsion angles of the peptide, 6 variables that define the position and orientation of the peptide ligand and the side-chain torsion angles of the flexible residues of the protein

**Table 2.** Summary of results from 'rigid' docking and 'induced fit' protein-peptide docking in MOLSDOCK.

|  | **Native** | **Best Sampled** | | | **Lowest Energy** | | | **Rank** |
|---|---|---|---|---|---|---|---|---|
|  | **Energy (no unit)** | **Energy (no unit)** | **RMSD (Å)** | **RDE (%)** | **Energy (no unit)** | **RMSD (Å)** | **RDE (%)** | **(%)** |
| Rigid | -353.29 | -326.44 | 1.61 | 43.96 | -400.95 | 2.66 | 55.36 | 2.5 |
| Induced fit | -870.81 | -859.77 | 1.62 | 46.78 | -949.43 | 1.69 | 53.11 | 4.2 |

The RMSD is calculated with respect to the native structure as described in the text. Rank (%) shows the ranking of the best sampled structure among the energy ranked structures.